\documentclass[12pt,letter]{article}

\usepackage{subfigure}
\usepackage{graphicx, epsfig, color}
\usepackage{amsmath,amssymb}
\def\to{\rightarrow}

\def\bi{\begin{itemize}}
\def\ei{\end{itemize}}

\def\sps1ap{SPS1a$^\prime$}
\def\c1p{C1$^\prime$}

\def\tst{\tilde t}

\def\tg{\tilde g}

\def\alt{\stackrel{<}{\sim}}
\def\agt{\stackrel{>}{\sim}}
\def\be{\begin{equation}}  
\def\ee{\end{equation}}  
\def\bea{\begin{eqnarray}}  
\def\eea{\end{eqnarray}}  
\def\beas{\begin{eqnarray*}}  
\def\eeas{\end{eqnarray*}}

\begin{document}
\begin{titlepage}
\begin{flushright}
OU-HEP-180930
\end{flushright}

\vspace{0.5cm}
\begin{center}
{\Large \bf Gravity safe, electroweak natural axionic\\
solution to strong $CP$ and SUSY $\mu$ problems
}\\ 
\vspace{1.2cm} \renewcommand{\thefootnote}{\fnsymbol{footnote}}
{\large Howard Baer$^{2}$\footnote[1]{Email: baer@nhn.ou.edu },
Vernon Barger$^{3}$\footnote[2]{Email: barger@pheno.wisc.edu } and
Dibyashree Sengupta$^{2}$\footnote[3]{Email: Dibyashree.Sengupta-1@ou.edu},
}\\ 
\vspace{1.2cm} \renewcommand{\thefootnote}{\arabic{footnote}}
{\it 
$^2$Homer L. Dodge Dep't of Physics and Astronomy,
University of Oklahoma, Norman, OK 73019, USA \\
}
{\it 
$^3$Dep't of Physics,
University of Wisconsin, Madison, WI 53706, USA \\
}
\end{center}

\vspace{0.5cm}
\begin{abstract}
\noindent 
Particle physics models with Peccei-Quinn (PQ) symmetry breaking 
as a consequence of supersymmetry (SUSY) breaking are attractive in that 
they solve the strong CP problem with a SUSY DFSZ-like axion, 
link the SUSY breaking and PQ breaking intermediate mass scales 
and can resolve the SUSY $\mu$ problem with a naturalness-required 
weak scale $\mu$  term whilst soft SUSY breaking terms inhabit the 
multi-TeV regime as required by LHC sparticle mass limits and 
the Higgs mass measurement. 
On the negative ledger, models based on global
symmetries suffer a generic gravity spoliation problem. 
We present two models based on the discrete $R$-symmetry 
${\bf Z}_{24}^R$-- which may emerge 
from compactification of 10-d Lorentzian spacetime in string theory-- 
where the $\mu$ term and dangerous proton decay and $R$-parity 
violating operators are either suppressed or forbidden 
while a gravity-safe PQ symmetry emerges as an accidental approximate 
global symmetry leading to a solution to the strong CP problem 
and a weak-scale/natural value for the $\mu$ term. 
\vspace*{0.8cm}


\end{abstract}

\end{titlepage}

While the discovery \cite{atlas_h,cms_h} of a very Standard Model (SM)-like 
Higgs boson at the CERN Large Hadron Collider (LHC) brings 
the experimentally established particle content into accord with the SM, 
there remain key reasons to expect that new physics {\it beyond the SM} 
will soon emerge.
Included among these are: 
1. the strong $CP$ problem in the QCD sector, 
2. the Higgs mass instability problem in the EW sector, 
3. the lack of a dark matter (DM) candidate and 
4. unification with gravity and the dark energy/cosmological constant problem.
The most elegant solution to the first of these is to introduce a
spontaneously broken global Peccei-Quinn (PQ) symmetry which leads to the 
presence of an axion\cite{pqww,ksvz,dfsz}. 
The axionic potential can dynamically settle to zero
thus eliminating the troublesome $CP$ violating gluon gluon-dual term 
in the QCD Lagrangian and as a bonus generate the axion which can serve as 
cold dark matter (CDM)\cite{axioncdm} and thus contribute in solving problem \#3 
as well. While the PQ solution to the strong CP problem is indeed compelling, 
unfortunately global symmetries are expected to be violated 
by the inclusion of gravity
into the theory\cite{grav} (black holes can swallow global charges) 
which then can spoil the PQ solution. PQ violating (PQV) terms in the
scalar potential must be suppressed by powers of the (reduced) 
Planck mass to at least the eighth power\cite{kmr}: 
$V_{PQV}\sim \phi^{12}/m_P^8$. This requirement for a gravity-safe
global PQ symmetry is enormously restrictive!
A further problem for axion models which may arise from string theory is that
the scale for PQ breaking-- characterized by the axion decay constant $f_a$--
is expected to occur around the grand unified mass scale 
$m_{GUT}\sim 10^{16}$ GeV\cite{witten}. 
Such a high value of $f_a$ typically leads to a vast 
overproduction of axion dark matter\cite{vg}. 
However, in models where PQ breaking is related to SUSY breaking 
at an intermediate mass scale, then instead $f_a$ typically arises 
within the range $10^{11}$--$10^{12}$ GeV 
and hence solves the problem of overproduction of axion cold dark matter.

Perhaps the most elegant solution to problem \#2 is to introduce weak scale
supersymmetry (SUSY)\cite{wss}. In the minimal supersymmetrized SM or MSSM,
the well-known Higgs mass quadratic divergences all cancel leaving only
mild log divergences to $m_h^2$. 
Indirect evidence for softly broken SUSY with weak scale superpartners
exists in that: 
1. The coupling strengths of the strong and electroweak
forces, as measured to high precision at the CERN LEP $e^+e^-$ collider
at energy scale $\sqrt{s}=m_Z$, enjoy an impressive unification at
$Q\simeq 2\times 10^{16}$ GeV when extrapolated to high energies\cite{gauge}. 
2. The top mass, measured
to be $m_t\simeq 173.2$ GeV, lies in the range required to trigger a
radiatively-induced breakdown of electroweak symmetry\cite{rewsb}. 
Finally, 3. the
light Higgs mass was found to lie at $m_h\simeq 125$ GeV, squarely
within the range required by the MSSM, 
where $m_h$ is bounded by $\alt 135$ GeV\cite{mhiggs}.

In spite of these impressive successes, a variety of problems arise in
SUSY theories-- foremost among these being the lack of appearance 
of the required superpartners at the CERN LHC.
Recent analyses of data from LHC run 2 with $\sqrt{s}=13$ TeV $pp$ 
collisions and $\sim 36$ fb$^{-1}$ of integrated luminosity 
require the gluino mass $m_{\tg}\agt 2$ TeV\cite{lhc_mgl} 
and the top squark mass $m_{\tst_1}\agt 1$ TeV\cite{lhc_mt1}.
Such large mass limits are far beyond initial expectations from naturalness
where for instance Barbieri-Giudice\cite{BG} (BG)-- requiring no worse than 
3\% electroweak finetuning-- expected that $m_{\tg}$ and $m_{\tst_1}$
are both $\alt 450$ GeV. 
Before declaring SUSY to be in a finetuning crisis, it was
pointed out in Ref's \cite{dew} that the BG bounds were computed in 
multiple soft parameter effective SUSY theories: in this case the BG 
calculation shows fine-tuning in the effective theory calculation but 
not in nature herself (as exemplified by more fundamental theories) 
wherein all soft parameters are 
interdependent and derived from more fundamental parameters 
(such as the gravitino mass $m_{3/2}$ in gravity-mediated SUSY breaking\cite{sugra} 
or the messenger scale $\Lambda$ in gauge mediation\cite{gmsb}).
For {\it correlated} soft parameters, then the EW fine-tuning may be 
extracted from the MSSM scalar potential minimization conditions 
which relate the measured $Z$-boson mass $m_Z$ to the 
{\it weak scale} SUSY Lagrangian parameters:
\bea 
\frac{m_Z^2}{2}& =& \frac{m_{H_d}^2 + \Sigma_d^d -(m_{H_u}^2+\Sigma_u^u)\tan^2\beta}{\tan^2\beta -1} -\mu^2\\ 
&\simeq & -m_{H_u}^2-\Sigma_u^u(\tst_{1,2})-\mu^2 .
\label{eq:mzs}
\eea 
Here, $m_{H_u}^2$ and $m_{H_d}^2$ are squared soft SUSY breaking
Lagrangian terms, $\mu$ is the superpotential Higgsino mass parameter,
$\tan\beta =v_u/v_d$ is the ratio of Higgs field
vacuum-expectation-values and the $\Sigma_u^u$ and $\Sigma_d^d$
contain an assortment of radiative corrections, the largest of which
typically arise from the top squarks.  Expressions for the $\Sigma_u^u$
and $\Sigma_d^d$ are given in the Appendix of Ref. \cite{rns}. 
The fine-tuning measure $\Delta_{EW}$ compares the largest independent 
contribution on the right-hand-side (RHS) of Eq.~(\ref{eq:mzs}) 
to the left-hand-side $m_Z^2/2$. 
If the RHS terms in Eq.~(\ref{eq:mzs}) are individually
comparable to $m_Z^2/2$, then no
unnatural fine-tunings are required to generate $m_Z=91.2$ GeV. 
The main requirements for low fine-tuning ($\Delta_{EW}\alt 30$\footnote{
The onset of fine-tuning for $\Delta_{EW}\agt 30$ is visually displayed in Ref. \cite{upper}.}) 
are the following.
\bi
\item $|\mu |\sim 100-350$
  GeV\cite{Chan:1997bi,kitnom,Barbieri:2009ew,hgsno,oldnsusy}
(where $\mu \agt 100$ GeV is required to accommodate LEP2 limits 
from chargino pair production searches).
\item $m_{H_u}^2$ is driven radiatively to small, and not large,
negative values at the weak scale~\cite{ltr,rns}.
\item The top squark contributions to the radiative corrections
$\Sigma_u^u(\tst_{1,2})$ are minimized for TeV-scale highly mixed top
squarks\cite{ltr}.  This latter condition also lifts the Higgs mass to
$m_h\sim 125$ GeV.  For $\Delta_{EW}\alt 30$, the lighter top
squarks are bounded by $m_{\tst_1}\alt 3$ TeV.
\item The gluino mass, which feeds into the $\Sigma_u^u(\tst_{1,2})$ via
renormalization group contributions to the stop masses\cite{oldnsusy}, 
is required to be $m_{\tg}\alt 6$ TeV, possibly beyond the reach of 
the $\sqrt{s}=13-14$ TeV LHC.\footnote{The upper bound on $m_{\tg}$ increases 
to 9 TeV for the natural anomaly-mediated SUSY breaking model\cite{namsb,lhc2}.}
\item First and second generation squark and slepton masses may range as
high as 5-30 TeV with little cost to naturalness\cite{cohen,rns,maren,upper}.
\ei 

In light of the finetuning clarification, 
it is perhaps not surprising that SUSY has yet to emerge at the LHC. 
Indeed, simple statistical arguments from the string theory landscape\cite{land}
suggest a pull to {\it large} values of soft terms albeit tempered
by the (anthropic) requirement that the weak scale not stray too far from its
value $m_{weak}\equiv m_{W,Z,h}\sim 100$ GeV. 
In this latter case, then present LHC mass limits are just beginning to 
probe natural SUSY parameter space and indeed it may require an energy 
upgrade of LHC to $\sqrt{s}\sim 27$ TeV for a full exploration\cite{lhc2}.

Along with non-appearance of superpartners, the MSSM suffers several \
structural problems arising from the superpotential. 
Including non-renormalizable terms (up to $m_P^{-1}$) 
as should be present in supergravity, 
then the gauge invariant MSSM superpotential reads:
\bea
W_{MSSM} &\ni &  \mu H_u H_d+\kappa_i L_i H_u+m_N^{ij}N^c_iN^c_j \label{eq:W} \\
 +f_e^{ij}L_iH_dE_j^c &+& f_d^{ij}Q_iH_dD_j^c+f_u^{ij}Q_iH_uU_j^c
+f_\nu^{ij}L_iH_uN^c_j\nonumber \\
+\lambda_{ijk}L_iL_jE_k^c & +&\lambda^{\prime}_{ijk}L_iQ_jD_k^c+\lambda_{ijk}^{\prime\prime}U_i^cD_j^cD_k^c\nonumber \\
&+& \frac{\kappa_{ijkl}^{(1)}}{m_P}Q_iQ_jQ_kL_l+
\frac{\kappa_{ijkl}^{(2)}}{m_P}U_i^cU_j^cD_k^cE_l^c .\nonumber
\eea
The first term on line 1 of Eq.~\ref{eq:W}, if unsuppressed, should lead to
Planck-scale values of $\mu$ while phenomenology (Eq.~\ref{eq:mzs}) 
requires $\mu$ of order the weak scale $\sim 100-350$ GeV. 
This is the famous SUSY $\mu$  problem albeit modified for the LHC era: 
why is $\mu\sim 100-350$ GeV whilst LHC Higgs mass and sparticle limits
require soft terms $m_{soft}\sim $multi-TeV? 
The $\kappa_i$, $\lambda_{ijk}$, $\lambda^{\prime}_{ijk}$ and 
$\lambda^{\prime\prime}_{ijk}$ terms violate either baryon number $B$ 
or lepton number $L$ or both 
and can, if unsuppressed, lead to rapid proton decay and an unstable 
lightest SUSY particle (LSP). 
The $f_{u,d,e}^{ij}$ are the quark and lepton Yukawa couplings and 
must be allowed to give the SM fermions mass via the Higgs mechanism.
The $\kappa_{ijkl}^{(1,2)}$ terms lead to dimension-five
proton decay operators and are required to be either 
highly suppressed or forbidden. 
It is common to implement discrete symmetries to forbid the 
offending terms and allow the required terms in \ref{eq:W}. 
For instance, the ${\bf Z}_2^M$ matter parity (or $R$-parity) forbids the
$\kappa_i$ and $\lambda_{ijk}^{(\prime ,\prime\prime)}$ terms but allows
for $\mu$ and the $\kappa_{ijkl}^{(1,2)}$ terms: 
thus, the ad-hoc $R$-parity conservation all by itself is 
insufficient to cure all of the ills of Eq. \ref{eq:W}. 

A promising approach, which addresses both the strong $CP$ problem and 
many of the offending terms in Eq. \ref{eq:W}, is to implement models 
where PQ symmetry breaking\cite{msy,cck,spm,primer} occurs 
as a consequence of SUSY breaking. 
In this approach, one posits the presence of a global PQ symmetry with 
PQ charges as listed in Table \ref{tab:PQ} along with a superpotential 
which includes the Yukawa couplings (second line) of Eq. \ref{eq:W} 
but then introduces additional chiral 
superfields $X$ and $Y$ and augments the superpotential with terms 
\be
W_{PQ}\ni \frac{1}{2}h_{ij}XN_i^cN_j^c +\frac{f}{m_P}X^3Y +W_\mu 
\label{eq:Wprime}
\ee
along with three possibilities for the bilinear $H_u H_d$ couplings:
\bea
W_\mu &\ni & \frac{g_{MSY}}{m_P} XYH_uH_d\ \ {\rm MSY\ model\cite{msy}} \label{eq:msy}\\
& & \frac{g_{CCK}}{m_P} X^2H_uH_d \ \ {\rm CCK\ model\cite{cck}} 
\label{eq:cck} \\
& & \frac{g_{SPM}}{m_P} Y^2 H_u H_d \ \ {\rm SPM\ model\cite{spm}}\label{eq:spm}
\eea
The model is postulated to hold just below the reduced Planck scale $m_P$.
The global $U(1)_{PQ}$ forbids the $\mu$ term, the RPV terms, the Majorana
neutrino mass term and the (last line) dangerous proton decay terms  of 
Eq. \ref{eq:W}. 
But when one augments the radiative PQ Lagrangian with 
soft SUSY breaking terms and allows for running of the PQ parameters 
$m_X^2$, $m_Y^2$, $h_{ij}$ and trilinears (for simplicity, we will here adopt
$h_{ij}=h_i\delta_{ij}$ as diagonal in generation space and assume
$h_1=h_2=h_3\equiv h$), then it is found-- for large PQ Yukawa coupling 
$h\sim 1.5-4$-- that $m_X^2$ is driven radiatively to negative values at
an intermediate scale resulting in spontaneous PQ symmetry breaking
wherein the $X$ and $Y$ fields develop vevs $v_X$ and $v_Y$ respectively.
The spontaneously broken global PQ symmetry generates a Goldstone boson--
the axion which solves the strong CP problem-- but then also generates
a superpotential mu term $\mu =\frac{g_{MSY}}{m_P}v_Xv_Y$, 
$\frac{g_{CCK}}{m_P}v_X^2$ or $\frac{g_{SPM}}{m_P}v_Y^2$ for the three
possibilities. The PQ scale $v_{PQ}=\sqrt{v_X^2+v_Y^2}$ and the
axion decay constant (given by $f_a=\sqrt{\sum_i q_{PQ}^2v_i^2}$)
depend on the SUSY PQ parameters via the scalar potential 
minimization conditions (listed {\it e.g.} in Ref. \cite{radpq}).

An attractive feature of the models-- 
similar to the Kim-Nilles model\cite{kn,bgw2}--
is that $\mu\sim f_a^2/m_P$ as compared to the soft SUSY breaking scale 
in gravity mediation $m_{soft}\sim m_{hidden}^2/m_P$ where $m_{hidden}$ is an
intermediate mass scale associated with the supergravity hidden sector. 
Thus, the aforementioned Little Hierarchy $\mu\ll m_{soft}$ 
ensues for $f_a< m_{hidden}$\cite{radpq}.
Also, since now the PQ scale $f_a$ is comparable to the hidden sector 
SUSY breaking scale, the expected axion dark matter density is generated in the 
cosmologically allowed range.
In addition, a Majorana neutrino scale is generated as 
$m_{N_i}=h_i v_X|_{Q=v_X}$ so that both of the intermediate 
PQ and Majorana neutrino
scales develop as a consequence of intermediate scale SUSY breaking.
\begin{table}[!htb]
\renewcommand{\arraystretch}{1.2}
\begin{center}
\begin{tabular}{c|ccccc}
multiplet & CCK & MSY & SPM & MBGW & GSPQ \\
\hline
$H_u$ & $-1$   & $-1$  & $-1$ & -1 & -1 \\
$H_d$ & $-1$   & $-1$  & $-1$ & -1 & -1 \\
$Q$   & $3/2$  & $1/2$ & $1/2$ & 1 & 1 \\
$L$   & $3/2$  & $1/2$ & $5/6$ & 1 & 1 \\
$U^c$ & $-1/2$ & $1/2$ & $1/2$ & 0 & 0 \\
$D^c$ & $-1/2$ & $1/2$ & $1/2$ & 0 & 0 \\
$E^c$ & $-1/2$ & $1/2$ & $1/6$  & 0 & 0 \\
$N^c$ & $-1/2$ & $1/2$ & $1/6$   & 0 & 0 \\
$X$   & $1$   & $-1$   & $-1/3$  & 1 & 1 \\
$Y$   & $-3$   & $3$   & $1$     & -1 & -3 \\
\hline
\end{tabular}
\caption{PQ charge assignments for various superfields of the CCK, MSY,
SPM, MBGW and GSPQ (hybrid CCK) models of PQ breaking from SUSY breaking. 
The gravity-safe hybrid SPM model will have the same PQ charges as 
GSPQ except $Q(X)=-1/3$ and $Q(Y)=1$.
}
\label{tab:PQ}
\end{center}
\end{table}

Unfortunately, these very appealing radiative PQ breaking scenarios 
are beset by the issue of the postulated global $U(1)_{PQ}$ symmetry suffering
from the previously mentioned gravity spoliation problem.
One way to deal with the gravity spoliation is to assume instead 
a gravity-safe discrete gauge symmetry ${\bf Z}_M$ of order $M$.
The ${\bf Z}_M$ discrete gauge symmetry can forbid the offending terms of
Eq. \ref{eq:W} while allowing the necessary terms\cite{chunlukas}.
Babu, Gogoladze and Wang\cite{bgw2} have found a closely related model 
(written previously by Martin\cite{spm} thus labelled MBGW) with
\be
W_{MBGW}\ni \lambda_{\mu}\frac{X^2H_uH_d}{m_P}+\lambda_2\frac{(XY)^2}{m_P}\label{eq:mbgw}
\ee
which is invariant under a ${\bf Z}_{22}$ discrete gauge symmetry.
These ${\bf Z}_{22}$ charge assignments have been shown to be anomaly-free 
under the presence of a Green-Schwarz (GS) term\cite{gs} 
in the anomaly cancellation calculation. The PQ symmetry then arises as
an {\it accidental approximate global symmetry} as a consequence 
of the more fundamental discrete gauge symmetry. The PQ charges of the
MBGW model are also listed in Table \ref{tab:PQ}.
The discrete gauge symmetry ${\bf Z}_M$ might arise if a charge $Me$ 
field condenses and is integrated out of the low energy theory while
charge $e$ fields survive (see Krauss and Wilczek, Ref.~\cite{grav}). 
While the ensuing low energy theory should be gravity safe, 
for the case at hand one might wonder at the 
plausibility of a condensation of a charge 22 object and whether it might 
occupy the so-called {\it swampland}\cite{swamp} of theories not 
consistent with a UV completion in string theory. 
In addition, the charge assignments\cite{bgw2} are not consistent with  
$SU(5)$ or $SO(10)$ grand unification which may be expected at some
level in a more ultimate theory.
Beside the terms in Eq.~\ref{eq:mbgw}, the lowest order 
PQV term in the superpotential is $\frac{(Y)^{11}}{m_P^8}$: 
thus this model is gravity safe. 

An alternative very compelling approach is to implement a 
discrete $R$ symmetry ${\bf Z}_N^R$ of order $N$.\footnote{
Discrete $R$ symmetries were used in regard to the $\mu$ problem in 
Ref.~\cite{choi_hall}
and for the PQ problem in Ref.~\cite{harigaya}. Accidental PQ symmetry from a
discrete flavor symmetry was examined recently in Ref.~\cite{Bjorkeroth:2017tsz}.} 
In fact, in Lee {\it et al.} Ref. \cite{lrrrssv1}, 
it was found that the requirement of an anomaly-free discrete symmetry that 
forbids the $\mu$ term and all dimenion four and five baryon and 
lepton number violating terms in Eq. \ref{eq:W} while allowing the 
Weinberg operator $LH_uLH_u$ and that commutes with $SO(10)$ 
(as is suggested by the unification of each family
into the 16 of $SO(10)$) has a unique solution: a ${\bf Z}_4^R$ $R$-symmetry.
If the requirement of commutation with $SO(10)$ is weakened to commutation with
$SU(5)$, then further discrete ${\bf Z}_N^R$ symmetries with
$N$ being an integral divisor of 24 are allowed\cite{lrrrssv2}: 
$N=4,6,8,12$ and 24. 
Even earlier\cite{bgw1}, the ${\bf Z}_4^R$ was found to be the simplest
discrete $R$-symmetry to realize $R$-parity conservation whilst 
forbidding the $\mu$ term. 
In that reference, the $\mu$ term was regenerated using 
Giudice-Masiero\cite{GM} which would generate $\mu\sim m_{soft}$ (too large).

$R$-symmetries are characterized by the fact that superspace co-ordinates 
$\theta$ carry non-trivial $R$-charge: 
in the simplest case, $Q_R(\theta )=+1$ so that $Q_R( d^2\theta ) =-2$. 
For the Lagrangian ${\cal L}\ni \int d^2\theta W$ to be invariant under 
$R$-symmetry, then the superpotential $W$ must carry $Q_R(W)= 2$. 
Discrete $R$ symmetries should be gravity-safe since they are expected to 
emerge as remnants of 10-$d$ Lorentz symmetry under compactification of 
extra dimensions in superstring theory.
The ${\bf Z}_N^R$ symmetry gives rise to a universal gauge 
anomaly $\rho$ mod $\eta$ where the remaining contribution $\rho$ 
is cancelled by the Green-Schwartz (GS) axio-dilaton shift and
$\eta =N$ ($N/2$) for $N$  odd (even).  
The anomaly free $R$ charges of various MSSM fields are listed in 
Table \ref{tab:R} for $N$ values consistent with grand unification.
\begin{table}[!htb]
\renewcommand{\arraystretch}{1.2}
\begin{center}
\begin{tabular}{c|ccccc}
multiplet & ${\bf Z}_{4}^R$ & ${\bf Z}_{6}^R$ & ${\bf Z}_{8}^R$ & ${\bf Z}_{12}^R$ & ${\bf Z}_{24}^R$ \\
\hline
$H_u$ & 0  & 4  & 0 & 4 & 16 \\
$H_d$ & 0  & 0  & 4 & 0 & 12 \\
$Q$   & 1  & 5 & 1 & 5  & 5 \\
$U^c$ & 1  & 5 & 1 & 5  & 5 \\
$E^c$ & 1  & 5 & 1 & 5  & 5 \\
$L$   & 1  & 3 & 5 & 9  & 9 \\
$D^c$ & 1  & 3 & 5 & 9  & 9 \\
$N^c$ & 1  & 1 & 5 & 1  & 1 \\
\hline
\end{tabular}
\caption{Derived MSSM field $R$ charge assignments for various anomaly-free 
discrete ${\bf Z}_{N}^R$ symmetries which are consistent with $SU(5)$ or 
$SO(10)$ unification (from Lee {\it et al.} Ref.~\cite{lrrrssv2}).
}
\label{tab:R}
\end{center}
\end{table}

We have examined whether or not the three radiative PQ breaking models of 
Table \ref{tab:PQ} (CCK, MSY and SPM) can be derived from any of the more
fundamental ${\bf Z}_N^R$ symmetries in Table \ref{tab:R}.
In almost all cases, the $hXN^cN^c$ operator is disallowed: then there is no
large Yukawa coupling present to drive the PQ soft term $m_X^2$ negative 
so that PQ symmetry is broken. And since the PQ symmetry does not allow for
a Majorana mass term $M_NN^cN^c$, then no see-saw scale can be developed.
One exception is the MSY model under ${\bf Z}_4^R$ symmetry with charge
assignments $Q_R(X)=0$ and $Q_R(Y)=2$: then a $YH_uH_d$ term is allowed
which would generate a $\mu$ term of order the intermediate scale.
Also, without considering any specific R-charges for the fields $X$ and $Y$, 
we can see that the R-charges for $X$ and $Y$ should be such that the term 
$XYH_uH_d$ is allowed and since the R-charges of $H_u$ and $H_d$ are 0, 
then a term  $MXY$ would always be allowed: this term breaks PQ 
at high order and is not gravity safe.
A second exception is SPM under the ${\bf Z}_6^R$ symmetry with
charges $Q_R(X)=0$ and $Q_R(Y)=2$: then operators like $Y^4/m_p$ are allowed
which break PQ but are not sufficiently suppressed so as to be gravity-safe. 
Furthermore, we can see that in this model that the R-charge of $Y$ 
is such that terms like  $M^2Y$ which break PQ are
always allowed but are not gravity safe.

We have also examined the MBGW model of Table~\ref{tab:PQ} which does allow 
for the $MN^cN^c$ see-saw term but where PQ and ${\bf Z}_N^R$ symmetry breaking is
triggered by large negative soft terms instead of radiative breaking.
To check gravity safety, we note that additional superpotential terms of the
form $\lambda_3X^pY^q$ may be allowed for given ${\bf Z}_N^R$ charge 
assignments and powers $p$ and $q$. Such terms will typically break the PQ
symmetry and render the model not gravity safe if scalar potential
terms $V(\phi )$ develop which are not suppressed by at least eight 
powers of $1/m_P$\cite{kmr}. 
The largest dangerous scalar potential terms develop
from interference between $\lambda_2 (XY)^2/m_P$ and 
$\lambda_3 X^pY^q/m_P^{p+q-3}$ when constructing the scalar potential
$V_F=\sum_{\hat{\phi}}|\partial W/\partial\hat{\phi}|_{\hat{\phi}\to\phi}$
(here, the $\hat{\phi}$ label chiral superfields with $\phi$ being their leading components).
We find the MBGW model to be not gravity safe under any of the ${\bf Z}_N^R$
discrete $R$-symmetries of Table \ref{tab:R}.

Next, we will adopt a hybrid approach between the radiative breaking models 
and the MBGW model by writing a superpotential:
\bea
W&\ni &f_uQH_uU^c+f_dQH_d D^c+f_{\ell}LH_dE^c+f_{\nu}LH_uN^c\nonumber \\
& +& fX^3Y/m_P+\lambda_\mu X^2 H_uH_d/m_P+M_NN^cN^c/2
\eea
along with PQ charge assignments given under the GSPQ (gravity-safe PQ model)
heading of Table \ref{tab:PQ}.
For this model, we have checked that there is gravity spoliation for
$N=4,\ 6,\ 8$ and 12. But for ${\bf Z}_{24}^R$ and under $R$-charge
assignments $Q_R(X)=-1$ and $Q_R(Y)=5$, then the lowest order
PQ violating superpotential operators allowed are 
$X^8Y^2/m_P^7$, $Y^{10}/m_P^7$ and $X^4Y^6/m_P^7$. 
These operators\footnote{The $X^8Y^2/m_P^7$ term was noted previously 
in Ref. \cite{lrrrssv2}.} 
lead to PQ breaking terms in the scalar potential suppressed by powers
of $(1/m_P)^8$. For instance, the term $X^8Y^2/m_P^7$  leads to
$V_{PQ}\ni 24f\lambda_3^*X^2YX^{*7}Y^{*2}/m_P^8+h.c.$ which is sufficiently 
suppressed by enough powers of $m_P$ so as to be gravity safe\cite{kmr}.
We have also checked that hybrid model using the MSY $XYH_uH_d/m_P$ 
term is not gravity-safe under any of the discrete $R$-symmetries of
Table \ref{tab:R} but the  
hybrid SPM model with $Y^2H_uH_d/m_P$ and charges $Q_R(X)=5$ and $Q_R(Y)=-13$ 
is gravity-safe under only ${\bf Z}_{24}^R$.

\begin{figure}[tbp]
\includegraphics[height=0.4\textheight]{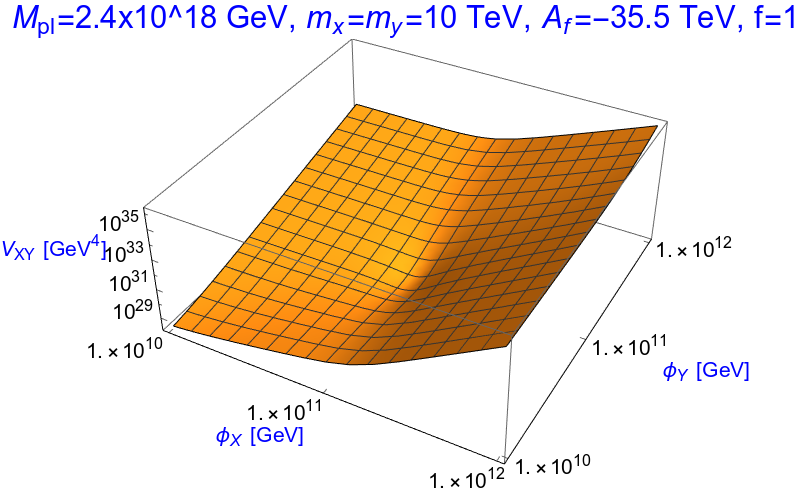}
\caption{Scalar potential $V_{GSPQ}$ versus $\phi_X$ and $\phi_Y$ for
$m_X=m_Y\equiv m_{3/2}=10$ TeV, $f=1$ and $A_f=-35.5$ TeV.
\label{fig:Vgspq}}
\end{figure}

We augment the hybrid CCK model scalar potential $V_F=|3f\phi_X^2\phi_Y/m_P|^2+|f\phi_X^3/m_P|^2$
of the GSPQ model by the following soft breaking terms
\be
V_{soft}\ni m_X^2|\phi_X|^2+m_Y^2|\phi_Y|^2+(f A_f\phi_X^3\phi_Y/m_P+h.c.)
\ee
and then minimize the resultant scalar potential. The minimization
conditions are already written down in Ref. \cite{radpq} Eq's 17-18.
In the case of the GSPQ model, the PQ symmetry isn't broken radiatively, but instead can be broken by adopting a sufficiently large negative value of $A_f$
(assuming real positive couplings for simplicity).
The scalar potential admits a non-zero minimum in the fields 
$\phi_X$ and $\phi_Y$ for $A_f<0$ as shown in Fig. \ref{fig:Vgspq} 
which is plotted for the case of $m_X=m_Y\equiv m_{3/2}=10$ TeV, 
$f=1$ and $A_f=-35.5$ TeV. For these values, we find 
$v_X=10^{11}$ GeV, $v_Y=5.8\times 10^{10}$ GeV, $v_{PQ}=1.15\times 10^{11}$ GeV
and $f_a=\sqrt{v_X^2+9v_Y^2}=2\times 10^{11}$ GeV. These sorts of numerical 
values lie within the mixed axion/higgsino dark matter sweet spot
of cosmologically allowed values and typically give dominant 
DFSZ axion CDM with $\sim 5-10\%$ WIMP dark matter\cite{bbc,dfsz2,axpaper}.  
Under these conditions, the model develops a $\mu$ parameter
$\mu =\lambda_\mu v_X^2/m_P$ and for a value $\lambda_\mu =0.036$ then we
obtain a natural value of the $\mu$ parameter at $150$ GeV.

\begin{figure}[tbp]
\includegraphics[height=0.4\textheight]{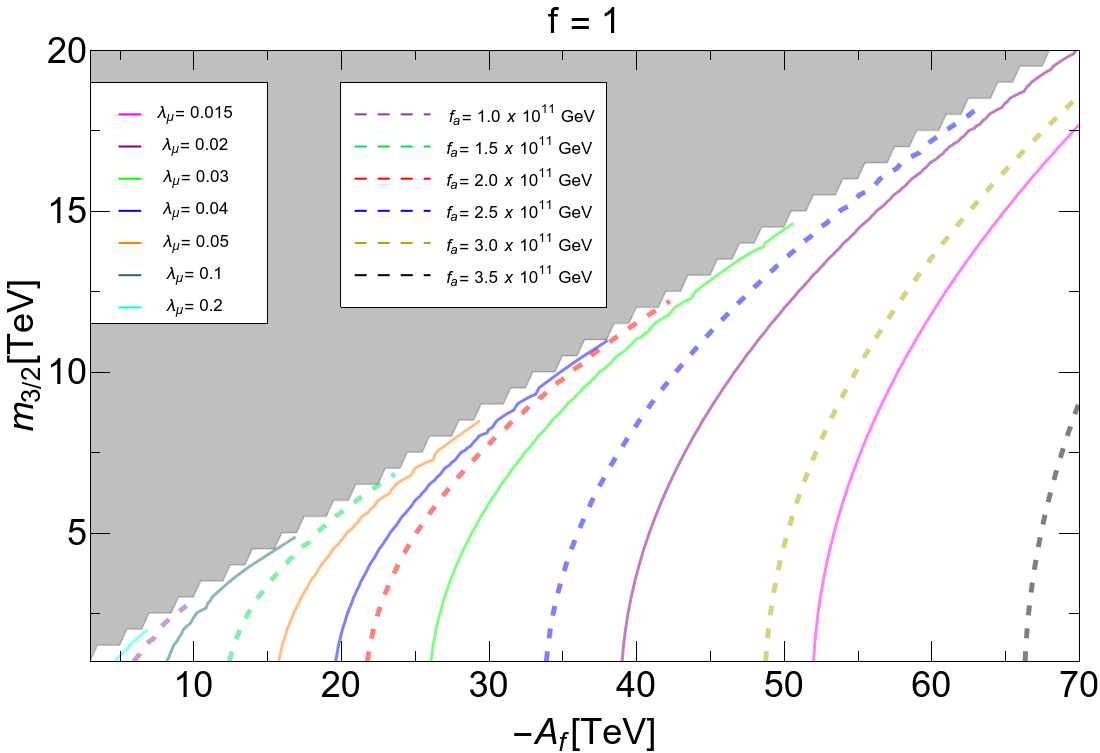}
\caption{Representative values of $\lambda_{\mu}$ required for $\mu =150$ GeV in the $m_{3/2}$ vs. $-A_f$ plane of the GSPQ model for $f=1$. 
We also show several contours of $f_a$.
\label{fig:GSPQ}}
\end{figure}

The allowed range of GSPQ model parameter space is shown in Fig. \ref{fig:GSPQ}
where we show contours of $\lambda_{\mu}$ values which lead to $\mu =150$ GeV
in the $m_{3/2}$ vs. $-A_f$ plane for $f=1$. We also show several
representative contours of $f_a$ values.
Values of $\lambda_{\mu}\sim 0.015-0.2$ are generally sufficient for a natural
$\mu$ term and are easily consistent with soft mass 
$m_{soft}\sim m_{3/2}\sim 2-30$ TeV as indicated by LHC searches.
We also note that for $m_{3/2}\sim 5-20$ TeV, then $f_a\sim 10^{11}$ GeV. 
Such high values
of $m_{3/2}$ also allow for a resolution of the early universe gravitino 
problem\cite{linde} (at higher masses gravitinos may decay before the onset of BBN)
and such high soft masses serve to ameliorate the SUSY flavor and CP 
problems as well\cite{masiero,dine}. 
They are also expected in several well-known
string phenomenology constructions including compactification of 
$M$-theory on a manifold of $G_2$ holonomy\cite{kane}, the minilandscape of
heterotic strings compactified on orbifolds\cite{mini} and the 
statistical analysis of the landscape of IIB intersecting $D$-brane 
models\cite{land}. 

As far as phenomenological consequences go, 
we may expect rather heavy sparticle masses 
which may require high-luminosity or high-energy LHC for 
verification of natural SUSY\cite{lhc2}.
The required light higgsinos with mass $\sim \mu$ may be observable at 
LHC\cite{bmt} or at a $\sqrt{s}\agt 500$ GeV $e^+e^-$ collider\cite{epem}.
The under-dense higgsino-like WIMP dark matter should eventually be detectable
at ton-scale noble liquid WIMP detectors\cite{wimp}. 
Regarding axion searches, the
axion anomaly contribution $E/N$ to the $a\gamma\gamma$ coupling\cite{Tanabashi:2018oca} is 
found (by including higgsino contributions to the anomaly diagram) 
to be $E/N=6/3$: this value is slightly larger than the chiral 
contribution to the $a\gamma\gamma$ coupling: 
together, the two contributions nearly cancel
leaving the axion visibility at microwave cavity experiments to be 
a very difficult prospect\cite{axion}.

{\it Summary:} We have found that the several 
DFSZ axionic extensions of the MSSM which generate PQ breaking 
radiatively as a consequence of SUSY breaking and a large PQ-neutrino Yukawa
coupling cannot be realized as a consequence of gravity-safe ${\bf Z}_N^R$
symmetries which are consistent with GUTs and forbid the $\mu$ term.
The MBGW model, where PQ breaking results from a large 
quartic soft term, also does not turn out to be gravity safe under 
${\bf Z}_N^R$ symmetries which are consistent with GUTs. 
However, the MBGW model does turn out to be gravity safe under  
${\bf Z}_{22}$ discrete gauge symmetry but this fundamental symmetry is
inconsistent with GUTs.
We have found (two) gravity-safe hybrid type models GSPQ 
with PQ superpotential as in the radiative models, 
but with an explicit see-saw neutrino sector
which is unrelated to SUSY or PQ breaking. 
Instead, the PQ breaking results as a consequence of a large quartic 
(Planck-suppressed) soft term so that it generates an axionic
solution to the strong CP problem along with a natural value of the 
MSSM $\mu$ term.
As emphasized in Ref. \cite{lrrrssv2}, the gravity-safe ${\bf Z}_{24}^R$
symmetry (which may emerge as a remnant of 10-$d$ Lorentz symmetry 
which is compactified to four spacetime dimensions)
yields an accidental approximate global PQ symmetry as implemented in the
GSPQ model of PQ symmetry breaking as a consequence of SUSY breaking. 
The ${\bf Z}_{24}^R$ (PQ) symmmetry breaking leads to 
$\mu\ll m_{soft}$ as required
by electroweak naturalness and to PQ energy scales 
$f_a\sim 10^{11}$ GeV as required by mixed axion-higgsino dark matter. 
The ${\bf Z}_{24}^R$ symmetry also forbids the dangerous dimension-four 
$R$-parity violating terms. 
Dimension-five proton decay 
operators are suppressed to levels well below experimental 
constraints\cite{lrrrssv2}.
Overall, our results show the axionic solution to the strong CP problem
is enhanced by the presence of both supersymmetry and 
extra spacetime dimensions which give rise to the gravity-safe 
${\bf Z}_{24}^R$ symmetry from which the required 
global PQ symmetry accidentally emerges.

\section*{Acknowledgments}

This work was supported in part by the US Department of Energy, Office of High Energy Physics.

%

%
\end{document}